\documentclass[runningheads]{llncs}
\usepackage[T1]{fontenc}
\usepackage{graphicx}
\begin{document}
\title{Initial Study On Improving Segmentation By Combining Preoperative CT And Intraoperative CBCT Using Synthetic Data}
\titlerunning{Initial Study on Multimodal Learning to Improve Segmentation}
%
%

\author{ Maximilian~E. Tschuchnig\inst{1,3}\and
Philipp Steininger\inst{2}\and
Michael Gadermayr\inst{1}}
\authorrunning{Tschuchnig et al.}
%

\institute{
Salzburg University of Applied Sciences, Applied Data Science Lab \and
MedPhoton GmbH \and
University of Salzburg, Department of Artificial Intelligence and Human Interfaces
}

\maketitle              
\begin{abstract}
Computer-Assisted Interventions enable clinicians to perform precise, minimally invasive procedures, often relying on advanced imaging methods. Cone-beam computed tomography (CBCT) can be used to facilitate computer-assisted interventions, despite often suffering from artifacts that pose challenges for accurate interpretation. While the degraded image quality can affect image analysis, the availability of high quality, preoperative scans offers potential for improvements. Here we consider a setting where preoperative CT and intraoperative CBCT scans are available, however, the alignment (registration) between the scans is imperfect to simulate a real world scenario. 
We propose a multimodal learning method that fuses roughly aligned CBCT and CT scans and investigate the effect on segmentation performance. For this experiment we use synthetically generated data containing real CT and synthetic CBCT volumes with corresponding voxel annotations. 
We show that this fusion setup improves segmentation performance in $18$ out of 20 setups investigated.
\end{abstract}
\section{Introduction}
\label{sec:introduction}
To establish computer-assisted interventions, precise and reliable imaging, especially intraoperative imaging, is crucial. Mobile robotic medical imaging systems, like cone-beam computed tomography (CBCT)~\cite{rafferty2006intraoperative}, enable intraoperative medical imaging with real time capabilities.
CBCT is an imaging method that utilizes a cone-shaped X-ray beam and a flat-panel detector to capture detailed, three-dimensional images of a patient's anatomy using a potentially mobile system~\cite{jaffray2002flat}.
However, using CBCT as an intraoperative imaging system often comes with the disadvantage of suffering from more artifacts than preoperative CT imaging, affecting the performance of downstream tasks like segmentation~\cite{wei2024reduction}.

While this degraded image quality can affect medical imaging tasks, the availability of high quality preoperative scans represents potential to integrate highly detailed information based on the idea of multimodal learning~\cite{zhang2021deep,podobnik2023multimodal,zhang2021modality}. Multimodal learning is an approach that involves fusing information from multiple domains to improve machine learning models for a downstream task like segmentation. In 3D medical imaging, a common approach is to enrich computed tomography (CT) scans, focusing on dense structures, with magnetic resonance imaging data for soft tissue analysis~\cite{zhang2021deep,podobnik2023multimodal}. Multimodal learning is typically separated into three fusion strategies~\cite{zhang2021deep,zhang2021modality}: early, late and hybrid.

Early-fusion combines images of different modalities before being processed by an image processing model. Typically, the two domains are fused along a dimension additional to the spacial volume dimensions and processed jointly~\cite{zhang2021deep,ren2021comparing}.
Late-fusion merges different streams before the final layer of the downstream task and hybrid fusion combines ideas of both early and late-fusion.

We propose a method that combines CBCT with misaligned, preoperative CT scans using early-fusion to improve semantic segmentation performance. Additionally, we examine how varying CBCT quality impacts segmentation performance in both baseline and multimodal configurations.

\section{Methodology}
We investigate multimodal learning of intraoperative CBCT volumes and misaligned, high quality, preoperative CT volumes using early-fusion, similar to Ren et al.~\cite{ren2021comparing} with both scans showing approximately the same section of the body. Evaluation is performed on the CBCT Liver Tumor Segmentation (CBCTLiTS)~\cite{tschuchnig2024cbctlits} dataset with the two different semantic segmentation targets, liver and liver tumor segmentation.

\begin{figure*}[!t]
\centerline{\includegraphics[width=\textwidth]{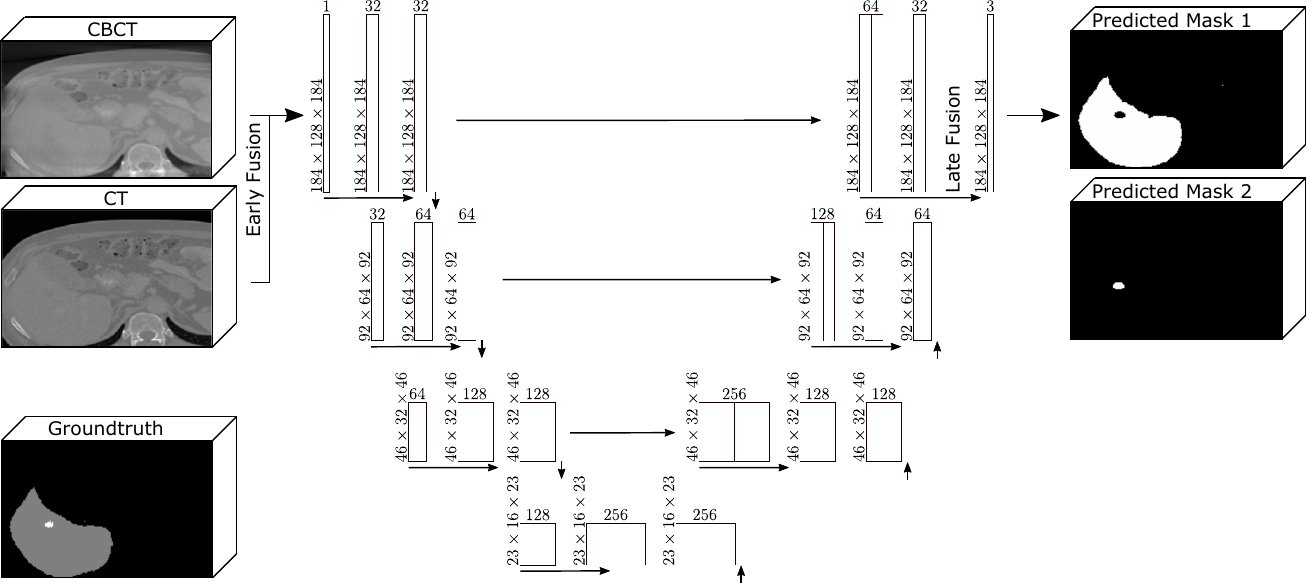}}
\caption{Multimodal model configuration. After fusing the intraoperative CBCT and preoperative CT (early fusion), the data was processed by the 3D unet, segmenting liver and liver tumors. The figure further shows where late fusion would be applied.}
\label{fig:model}
\end{figure*}

To perform segmentation we used a holistic 3D unet introduced by Çiçek et al.~\cite{cciccek20163d} as the basis of our segmentation model and our baseline for all investigated settings. The 3D unet was adapted to process multimodal data by adding a paired and misaligned, preoperative CT as a second channel, resulting in a 4d data structure (early  fusion).
The 3D unet used for segmentation is shown in Fig.~\ref{fig:model} and consists of an encoder with 3 double convolution layers and $3 \times 3 \times 3$ convolutional kernels, connected by 3D max pooling. The latent space consists of one double convolution block followed by the unet decoder, mirroring the encoder. As is typical for unet, each double convolutional output in the encoder is also connected to the decoder double convolutional block of the same order. Additionally, one 3D convolutional layer is added to the decoder with a filter size of $1 \times 1 \times 1$ and the number of filters set to the number of segmentation classes. The number of feature maps were set to $\{32, 64, 128, 256\}$. Batch norm was applied after each layer in the double convolutional blocks. The model was trained utilizing a sum of binary cross-entropy and Dice similarity. For our baseline, the same unet was used~\cite{cciccek20163d}, with only the CBCT as input.

\subsection{Data}
\begin{figure}[!t]
\includegraphics[width=\columnwidth]{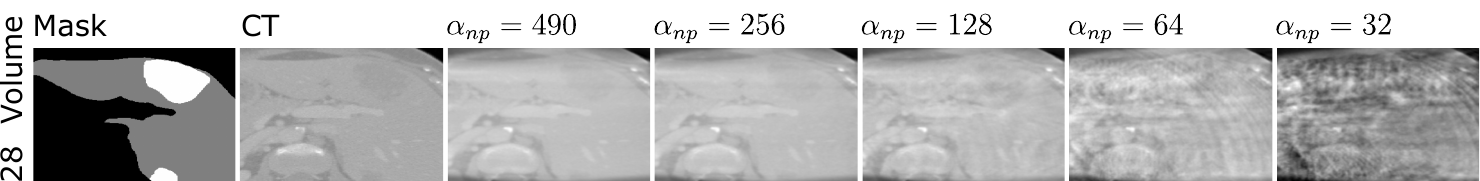}
\caption{Sample CBCTLiTS data (subject 28): From left to right, the segmentation mask, CT scan, and various CBCT scans are shown, with CBCT quality degraded according to the undersampling factor $\alpha_{np}$.} \label{fig:data}
\end{figure}

CBCTLiTS is an addition to the well studied Liver Tumor Segmentation (LiTS)~\cite{bilic2023liver} dataset. LiTS is an abdominal CT dataset with liver and liver tumor segmentations. CBCTLiTS~\cite{tschuchnig2024cbctlits} expands on that by adding perfectly aligned, synthetic CBCT volumes. Fig~\ref{fig:data} shows exemplar CT and corresponding CBCT volumes (with the varying parameter $\alpha_{np}$) of subject $28$, and the provided segmentation mask.

The factor $\alpha_{np}$ corresponds to the undersampling of Digitally Reconstructed Radiographs (DRRs) in the CBCT reconstruction process of CBCTLiTS. In detail, Tschuchnig et al.~\cite{tschuchnig2024cbctlits} synthetically generated CBCTLiTS by simulating Digitally Reconstructed Radiographs (DRRs) from the LiTS CT volumes, followed by filtered backprojection to generate the CBCT scans in CBCT geometry.
By adjusting undersampling based on the parameter $\alpha_{np} \in \{32, 64, 128, 256, 490\}$, the quality of the synthetic CBCT can be adapted. 

\begin{figure}[!t]
\includegraphics[width=1\columnwidth]{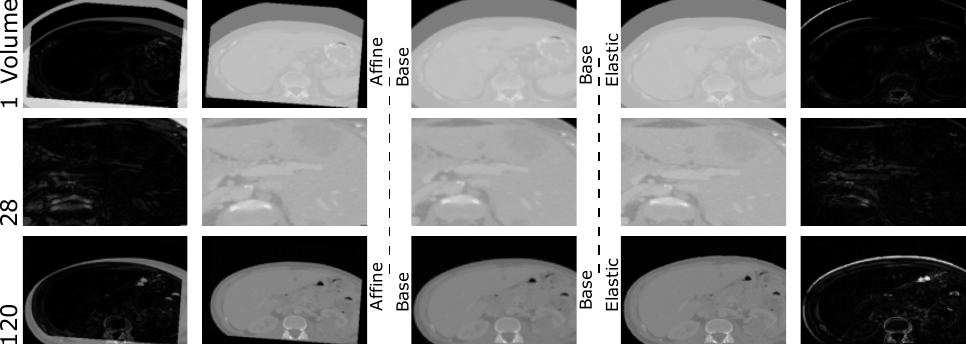}
\caption{On the left, results of random affine augmentation ($\alpha_a = 0.25$) and the corresponding absolute difference image of $3$ different CBCTLiTS volumes are shown. The centered column shows the original data. Right, resulting volumes of random elastic misalignment ($\alpha_a = 0.25$) of the same volumes and the corresponding difference images are shown. Elastic misalignment is shown separate from affine misalignment for easier readability.} \label{fig:augresultingdata}
\end{figure}

Since we aim to generate a realistic evaluation scenario, we synthetically generate misalignment between the perfectly aligned CT and CBCT pairs. We further aim to reduce the number of parameters controlling misalignment to one. Misalignment is established by the factor $\alpha_a$. Misalignment augmentation was performed using TorchIO RandomAffine and RandomElasticDeformation.
In detail, affine misaligned was performed using random (non-isotropic) scaling, with the scaling parameter sampled from $\mathcal{U}(1 - 0.5 \cdot \alpha_a,1 + 0.5 \cdot \alpha_a)$, rotation, parameters sampled from $\mathcal{U}(-22.5 \cdot \alpha_a, 22.5 \cdot \alpha_a)$, and translation, with the parameter sampled from $\mathcal{U}(0, 0.5 \cdot \alpha_a)$ with tri-linear interpolation. Elastic misalignment was applied with a maximum displacement sampled from $\mathcal{U}(0, 20 \cdot \alpha_a)$ with $7$ control points and no locked borders. Figure~\ref{fig:augresultingdata} shows $3$ unprocessed volumes and examples of affine and elastic misalignment with $\alpha_a = 0.25$. Figure.~\ref{fig:augresultingdata} displays the elastic transformation in isolation from affine transformations for improved readability.

During training and validation, only affine transformations were used to streamline data loading. In testing, both affine and elastic transformations were applied for a single evaluation per sample for a more realistic evaluation. The CBCTLiTS training set contains 131 paired samples. Since the CBCTLiTS testing data lacks segmentation masks, it was excluded from this study.

\subsection{Experimental Details}
The models were trained on a Ubuntu server using NVIDIA RTX A6000 graphics cards. Due to the large data size and 48 GB VRAM memory limit, volumes were downscaled (isotropic) by the factor of two~\cite{tschuchnig2024multi}. 
To binarize the masks, a threshold of $0.5$ was applied to each channel of the unet output. The data was separated into training-validation-testing data. The separation was performed using the ratios $0.7$ (training), $0.2$ (validation), $0.1$ (testing). All experiments were trained and evaluated $4$ times to facilitate stable results with the same random splits as well as the same random CT misaligned for comparable result. Adam was used as an optimizer with a learning rate of $0.005$ for liver and liver tumor segmentation

\section{Results}
\begin{figure}[!t]
\includegraphics[width=\columnwidth]{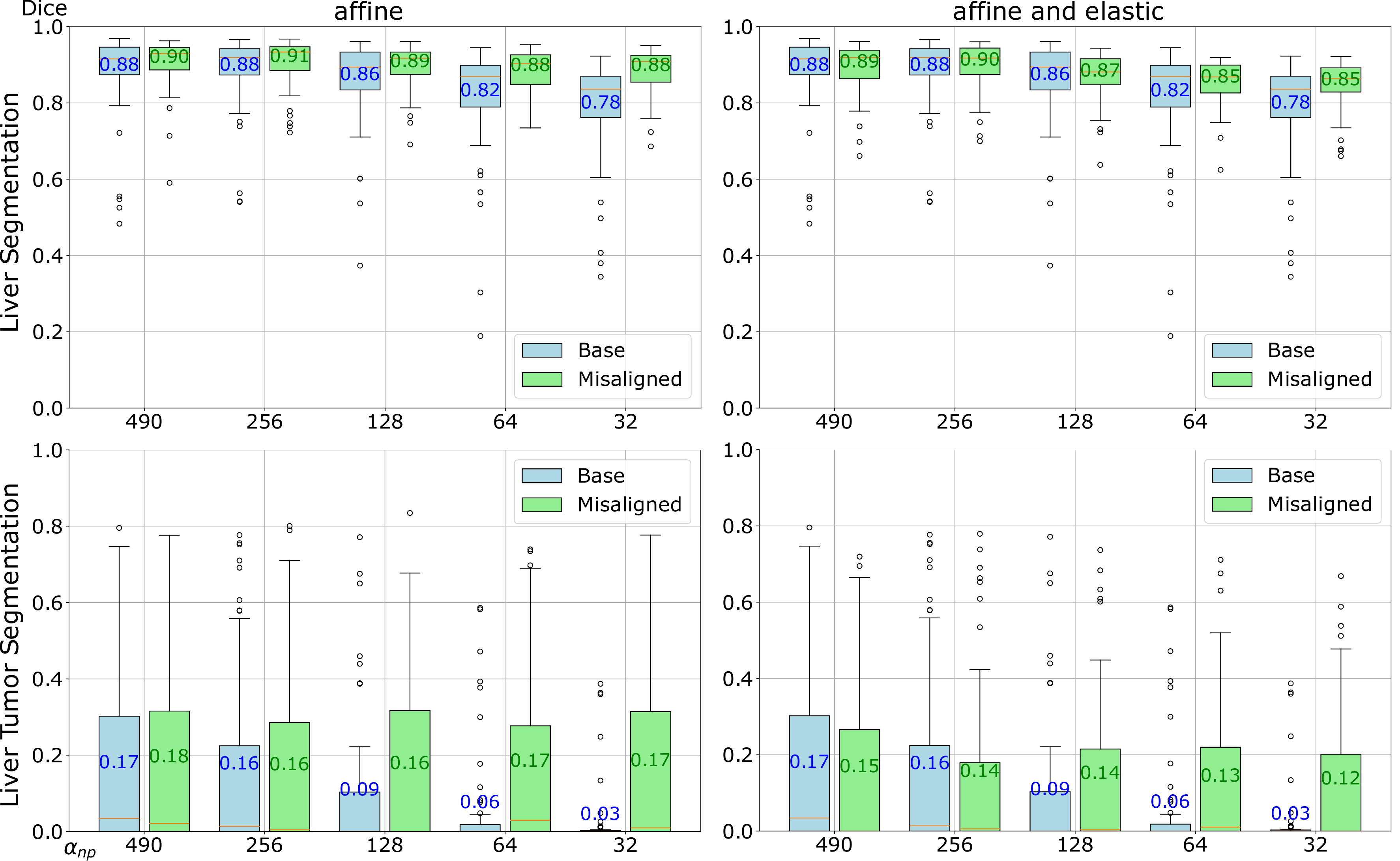}
\caption{Experimental results shown as boxplots. In the top row, results for liver segmentation are shown with Dice score on the y axis and the volume quality ($\alpha_{np}$) on the x axis. The median is shown as an orange bar and the mean value is printed in the corresponding box. Baseline results are displayed in blue and our proposed, multimodal method in green. The bottom row shows liver tumor segmentation. In the first column, experiments were performed on affine and the second column on combined affine and elastic transformed volumes.} \label{fig:results}
\end{figure}

Experimental results are shown as boxplots in Fig.~\ref{fig:results}. The top row shows the results of liver and the bottom row of liver tumor segmentation. The left column shows results evaluated on affine misalignment, while the right columns show combined affine and elastic misalignment. The undersampling factor $\alpha_{np}$ is shown on the x axis and Dice similarity coefficients on the y axis. Blue boxes show Baseline results and green boxes the multimodal learning approach results. Additionally the mean values are plotted in the corresponding box, for easy comparison.

\section{Discussion}
The results show that enriching intraoperative CBCT with roughly aligned, preoperative CT improves medical imaging tasks like semantic segmentation. Most multimodal setups improved segmentation performance with the only exception in liver tumor segmentation with $\alpha_{np} = 490$ in the combined affine and elastic misaligned case.

Several trends are notable. First, the worse the CBCT quality, the more can be gained by adding high quality CT, leading to a maximum increase of Dice from $0.78$ to $0.88$ for liver and from $0.03$ to $0.17$ for liver tumor segmentation. This shows that adding the preoperative CT partly mitigates the decreasing image quality of the intraoperative CBCT. This also shows that there might be some implicit registration taking place, however, further experimentation with different degrees of misalignment are needed to investigate this effect further.

Since we  generate a dynamic dataset for both training and testing through the misalignment process, the misalignment itself can be seen as data augmentation. We assume that this effect can further improve the positive effect of the proposed multimodal setup however, this has to be evaluated on real, imperfectly aligned volume pairs. Furthermore, this form of multimodal learning is theoretically applicable to other medical imaging applications and architectures. Due to the positive results of this study, further experimentation into other architectures, such as the Segment Anything Model~\cite{kirillov2023segment} or UNETR~\cite{hatamizadeh2022unetr} present high potential.

\begin{credits}
\subsubsection{\ackname} This project was partly funded by the Austrian Research Promotion Agency (FFG) under the bridge project "CIRCUIT: Towards Comprehensive CBCT Imaging Pipelines for Real-time Acquisition, Analysis, Interaction and Visualization" (CIRCUIT), no. 41545455 and by the county of Salzburg under the project AIBIA and the Salzburg University of Applied Sciences under the project FHS-trampoline-8 (Applied Data Science Lab).
\end{credits}
%
%
%
\bibliographystyle{splncs04}
\bibliography{bib}
\end{document}